\documentclass[twocolumn,showpacs,preprintnumbers,amsmath,amssymb]{revtex4}

\input epsf

\def\be{\begin{equation}}
\def\ee{\end{equation}}
\def\bea{\begin{eqnarray}}
\def\eea{\end{eqnarray}}

\def\keV{\,{\rm keV}}
\def\MeV{\,{\rm MeV}}

\def\kpc{\,{\rm kpc}}
\def\Mpc{\,{\rm Mpc}}

\def\eV{{\,\rm eV}}

\def\cmm2{{\,\rm cm^{-2}}}
\def\cm2{{\,{\rm cm}^2}}
\def\cmm3{{\,{\rm cm}^{-3}}}
\def\gcmm3{{\,{\rm g\,cm^{-3}}}}

\def\fun#1#2{\lower3.6pt\vbox{\baselineskip0pt\lineskip.9pt
  \ialign{$\mathsurround=0pt#1\hfil##\hfil$\crcr#2\crcr\sim\crcr}}}

\def\kpc{{\rm~kpc}}

\def\p3m{P$^3$M}

\def\fun#1#2{\lower3.6pt\vbox{\baselineskip0pt\lineskip.9pt
  \ialign{$\mathsurround=0pt#1\hfil##\hfil$\crcr#2\crcr\sim\crcr}}}

\newcommand{\Ms}{\,\mathrm{M_\odot}} 					
\newcommand{\arcmin}{\,\mathrm{arcmin}} 			
\newcommand{\arcsec}{\,\mathrm{arcsec}} 			
\newcommand{\ergcms}{\,\mathrm{erg/cm^2/sec}}	

\newcommand\eqnref[1]{%
  eq.~\ref{eqn:#1}}

\newcommand\figref[1]{%
  \figurename ~\ref{fig:#1}}

\begin{document}
\bibliographystyle{prsty}
\title{Probing the nature of dark matter with Cosmic X-rays:\\ Constraints from ``Dark blobs'' and grating spectra of galaxy clusters}
\author{Signe Riemer-Sorensen$^1$, Kristian Pedersen$^1$, Steen H. Hansen$^1$,
and Haakon Dahle$^{2}$}
\affiliation{$^1$ Dark Cosmology Centre, Niels Bohr Institute, University of Copenhagen, Juliane Maries Vej 30, DK-2100 Copenhagen, Denmark \\$^2$ Institute of Theoretical Astrophysics, University of Oslo, P.O. Box 1029, Blindern, N-0315 Oslo, Norway}
\date{\today}

\begin{abstract}
Gravitational lensing observations of galaxy clusters have identified
dark matter ``blobs'' with remarkably low baryonic content. 
We use such a system to probe the particle nature of
dark matter with X-ray observations. We also study high resolution
X-ray grating spectra of a cluster of galaxies.  From these grating
spectra we improve the conservative constraints on a particular dark matter candidate,
the sterile neutrino, by more than one order of magnitude. Based on these
conservative constraints obtained from Cosmic
X-ray observations alone, the low mass ($m_s \lesssim 10$~keV) and low
mixing angle ($sin^2(2\theta) \lesssim 10^{-6}$) sterile neutrino is still
a viable dark matter candidate.
\end{abstract}
\maketitle

\section{Introduction}

The cosmological dark matter abundance is firmly established through
observations of the cosmic microwave background and of the 
large scale structure of the universe \cite{spergel,uros}. 
This is complemented by measurements of dark matter
on smaller scales by studies of, e.g. the rotation curves of galaxies, 
gravitational lensing by galaxies and clusters of galaxies \cite{lensing}, the 
velocity dispersion of galaxies in clusters of galaxies, and X-ray
emitting hot gas in clusters of galaxies \cite{xraycluster}. However, the
particle nature of dark matter remains a puzzle.

There are numerous dark matter candidates, among which the sterile
neutrino is particularly well motivated. The sterile neutrino is a natural dark 
matter candidate in a minimally extended standard model of particle physics
\cite{dodelson:1993} and it provides solutions to
other problems: the masses of the active neutrinos \cite{asaka:2005a},
the baryon asymmetry of the Universe \cite{asaka:2005b}, and the
observed peculiar velocities of pulsars
\cite{kusenko:1997,fuller:2003}. Sterile neutrinos participate in the flavour-mass eigenstate
oscillations of the active standard model neutrinos, and are thereby
allowed to decay radiatively through a two-body decay with photon
energy predicted to lie in the X-ray range ($E_\gamma=m_s/2$, where $m_s$
is the rest mass of the sterile neutrino). This
renders it a testable dark matter candidate \cite{dolgovhansen}.

The decay rate of any dark matter candidate with a radiative two-body
decay can be constrained from observations of dark matter
concentrations (for references see \cite{boyarsky:2006b, savvas, watson:2006}). 
The strongest constraints are obtained from studying dark matter dominated 
regions, and with instruments with high spectral resolution (since the decay line
is expected only to suffer negligible broadening).  

Recent gravitational lensing observations of the mass distribution in galaxy clusters \cite{bullet:2006} have identified cluster scale dark matter ``blobs'' with very low baryonic content. This allows for the novel possibility of using such almost pure dark matter blobs to probe the particle nature of the dark matter \cite{hansen:2002}. Below we analyze X-ray observations of the dark matter blob in the cluster of galaxies Abell~520. Also, we analyze high resolution X-ray grating spectra of the cluster of galaxies Abell~1835, allowing us to improve constraints from earlier studies by more than one order of magnitude.

\section{X-ray Data Analysis}
When a spectrum has been obtained from an observation of a given dark matter dense region, there are different ways of searching for a hypothetical mono-energetic emission line and to determine an upper limit on the flux from decaying dark matter particles. The simplest and most conservative method is the ``slice method,'' where the energy range of the spectrum is divided into bins of a width equal to the instrumental energy resolution $(2\sigma)$, and all of the X-ray flux in a particular bin is determined from a model fitted to the spectrum. The slice method is very robust, as the physics behind the fitted model is irrelevant, and the method does not require any assumptions about the X-ray background, but regards all received flux as an upper limit for the flux from decaying dark matter. This is despite the fact that the total flux is known to consist of several contributions; the cosmic X-ray background from unresolved sources, the X-ray emission from the intra cluster medium, the Milky Way halo, and the instrumental background. The ``slice method'' takes into account that an emission line from decaying dark matter could ``hide'' under a line feature in the spectrum \cite{blanksky}. Other methods for constraining the flux (for example \cite{boyarsky:2006a}) can give stronger, but less robust, results (for a discussion of different methods see \cite{thesis}). In this study, we conservatively use the slice method.

\section{Grating Observations of The Galaxy Cluster Abell~1835} \label{sec:analysis}
A good spectral resolution is required to search for a mono-energetic emission line. The high spatial resolution of the {\it Chandra} X-ray telescope can be turned into a very high spectral resolution ($\approx 5 \eV$) by deflecting the incoming photons in a grating, as the deflection angle is highly sensitive to the photon energy. The ACIS-HETG instrument consists of two gratings, which on demand can be placed between the mirrors and the ACIS CCDs: the High Energy Grating (0.8--9.0\keV), HEG, and the Medium Energy Grating, MEG (0.4--5.0~keV) \footnote{http://cxc.harvard.edu/proposer/POG/html, http://space.mit.edu/CSR/hetg\_info.html}.

When the incoming photons are deflected in a grating, the information of their spatial origin is lost. This makes it impossible to optimize the ratio of expected dark matter signal to noise from X-ray emitting baryons in the observational field of view (as described by \cite{blanksky, watson:2006}). In this study, the cluster of galaxies Abell~1835 was targeted because most of the emission from the cluster comes from a region close to the observational axis making grating spectroscopy possible. Abell~1835 has a luminosity distance of $D_L=1225\Mpc$ ($z=0.252$, \cite{Schmidt:2001}), and the mass within the scale radius of $R \approx 800 \kpc \approx 4.2\arcmin$ is $M_{fov}^{A1835}=6.5\times10^{14} \Ms$ \cite{voigt:2006}. No other obvious mass concentrations are seen in the field of view. (Here, and in the following, a cosmology with $\Omega_m=0.26$, $\Omega_\Lambda=0.74$, $h=0.71$ is assumed).

The spectral resolution of a grating spectrometer decreases proportionally to the angular extension of the source. For Abell~1835 the emission is dominated by photons from within the core radius, $r_c=39\pm3 \kpc\approx 10\arcsec$ \cite{Schmidt:2001} so this characteristic radius was used to determine the spectral resolution \footnote{http://cxc.harvard.edu/proposer/POG/html}:
\begin{eqnarray} \label{eqn:sigma}
\sigma_{HEG} \approx \frac{(E_\gamma/\keV)^2}{156 m}\,\keV \, \\
\nonumber \sigma_{MEG} \approx \frac{(E_\gamma/\keV)^2}{78m}\,\keV \, ,
\end{eqnarray}
where $m$ is the dispersion order. Grating spectra of Abell~1835 were extracted from the observation with id 511 using CIAO 3.3 \footnote{http://cxc.harvard.edu/ciao}. The first order deflections were combined into a single spectrum with a spectral bin size of $0.006\keV$ and $0.013\keV$ (at $E_\gamma=1\keV$) for HEG and MEG, respectively. The line broadening due to the velocity dispersion of dark matter in Abell~1835 is negligible ($v/c\approx10^{-5}-10^{-4}$).

With the spectral fitting package Sherpa \cite{sherpa}, a thermal plasma model (MEKAL, \cite{mekal}) was fitted to the data with the temperature and abundance as free parameters. The model is an adequate representation of the data, giving a reduced $\chi^2$ of 1.2 in both cases (683 and 569 degrees of freedom for HEG and MEG, respectively). The flux was determined using the slice method for a slice width given by \eqnref{sigma}.

\section{The Dark Matter Blob in the Galaxy Cluster Abell~520}
\begin{figure}
\center
\epsfxsize=8.6cm
\epsfysize=8.5cm
\epsffile{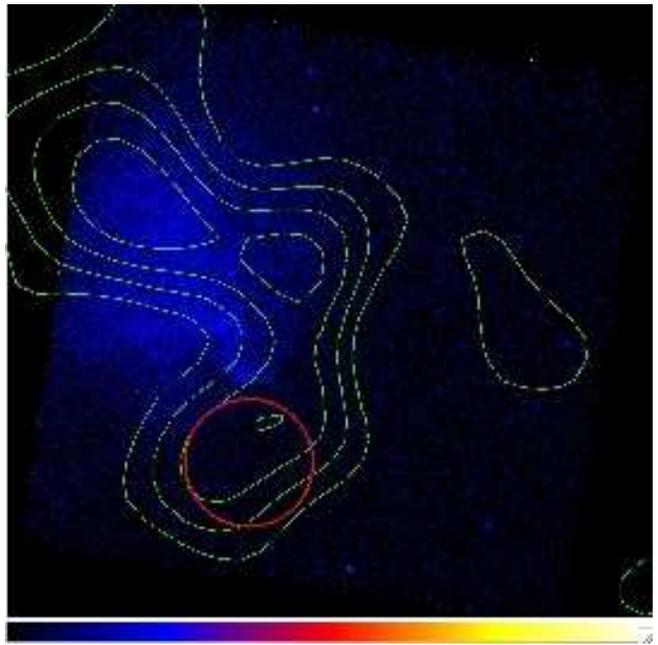}
\caption{(Color online) Abell~520 observed in X-rays (0.3-10.0~keV, blue color) with {\it Chandra} with the gravitational potential from weak lensing overlaid (green contours). The dark matter blob in the red circle has very low X-ray emission from baryons.}
\label{fig:A520}
\end{figure}

For direct imaging data the field of view can be optimized by observing a dark matter dense region with low X-ray emission from baryons. A unique example is the merging cluster of galaxies Abell~520 containing a ``blob'' of high mass concentration with very low X-ray emission discovered recently using weak gravitational lensing \cite{markevitch:2004}, see \figref{A520} \footnote{A similar analysis of the dark matter concentration in the bullet cluster has been carried out by Boyarsky et al. Their findings are qualitatively similar to the blob results presented here (private communication with M. Shaposhnikov)}.

An ACIS-S3 0.3--9.0~keV spectrum was extracted from the {\it Chandra} observation with id 4215 for a region centred at the dark blob ($RA,DEC=04^{\mathrm{h}} 54^{\mathrm{min}} 04.036^{\mathrm{s}},+02^{\circ} 52^{'} 37.30^{''}$) and with a radius of $r=0.85\arcmin = 190 \kpc$ (the red circle in \figref{A520}). A model consisting of a power law and six Gaussians was fitted to the spectrum with a reduced $\chi^2$ of 1.2 (for 80 degrees of freedom). The flux was determined using the slice method for a slice width given by the resolution of ACIS-S3 \footnote{http://cxc.harvard.edu/proposer/POG/html}:
\begin{equation}
\sigma_{S3}=0.005 E_\gamma + 0.05 \keV
\end{equation}

The mass of the dark matter blob in A520 has been derived from weak gravitational lensing to be {\bf $M^{blob}_{fov}=4.78\pm1.5\times10^{13}h^{-1}\Ms$}. This value is based on measuring the overdensity in the blob region
with respect to the mean density in a surrounding annulus with inner and outer radius of 0.85 arcmin and 4 arcmin, respectively. Hence, the mass value can be regarded as a conservative lower limit on the mass contained within the blob region. A detailed description of the data and methodology of the weak lensing analysis is given elsewhere \cite{dahle:2002}. 

The luminosity distance to A520 is $D_L=980\Mpc$ ($z=0.203$, \cite{ebeling:1998}).

\section{Decay Rate} \label{decayrate}
Let us now consider a specific dark matter candidate, the sterile
neutrinos. They can decay radiatively as $\nu_s \rightarrow \nu_\alpha +
\gamma$, where $\nu_\alpha$ is an active neutrino. This is a two-body
decay with a photon energy of $E_\gamma=m_s/2$. Assuming only one kind of dark matter, the observed flux,
$F_{obs}$, at a given photon energy yields an upper limit on the flux
from two-body radiatively decaying dark matter:
\begin{equation} \label{eqn:maxdecay}
\Gamma _{\gamma} \leq \frac{8 \pi F D_L^2}{M_{fov}}  \, .
\end{equation}

The Milky Way dark matter halo will always be included in the observation, and its mass contribution to the total mass in the field of view has to be taken into account \cite{blanksky, boyarsky:2006b}. For the Milky Way we use a virial mass of $M_{halo}^{vir} = 10^{12} \, \Ms$ and by integration of a NFW profile we find the halo mass to have a mean distance of $35\kpc$. The halo mass and mean distance varies less than a factor of two for a reasonable range of model parameters \cite{blanksky}. For the grating observation of Abell~1835 the Milky Way halo mass within the field of view is $M_{halo}^{A1835}=9\times10^{5}\Ms$, and for the dark matter blob in Abell~520 it is $M_{halo}^{A520}=4\times10^{3}\Ms$.

\begin{figure}
\center
\epsfxsize=8.6cm
\epsfysize=6.1cm
\epsffile{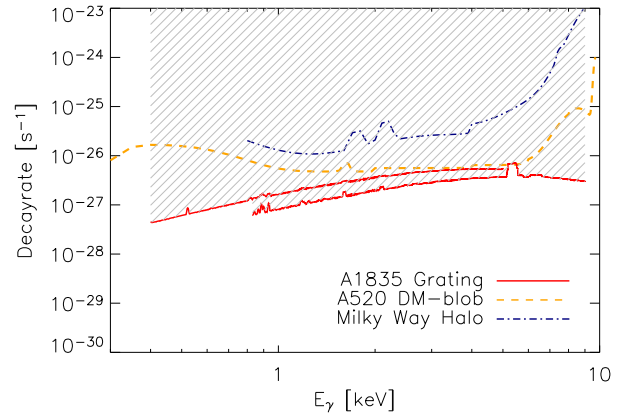}
\caption{(Color online) The upper limit on the radiative two-body decay rate obtained from the dark matter blob of Abell~520 (dashed) and the grating spectra of Abell~1835 (solid) shown together with the Milky Way halo constraint \cite{blanksky} (dot-dashed).}
\label{fig:decayrate}
\end{figure}

\figref{decayrate} shows the upper limit on the decay rate of any dark matter candidate (given by \eqnref{maxdecay}) obtained from the total amount of received flux for the Abell~520 dark matter blob and the Abell~1835 grating data. It is seen that the grating data of Abell~1835 provide constraints that are one to two orders of magnitude stronger than the constraints obtained from observations of the Milky Way halo alone \cite{blanksky,boyarsky:2006a}. Also, the data from the blob of Abell~520 provide stronger constraints than the Milky Way halo.

\begin{figure}
\center
\epsfxsize=8.6cm
\epsfysize=6.1cm
\epsffile{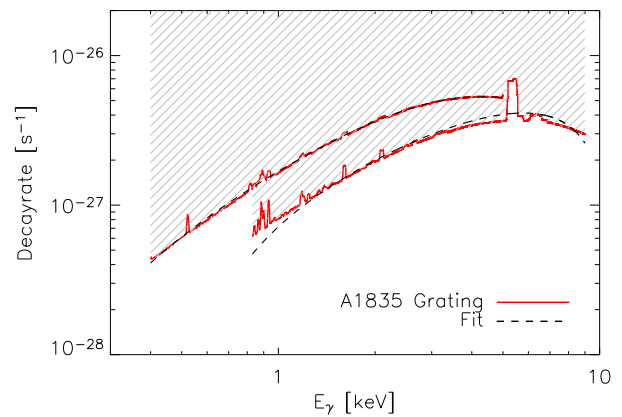}
\caption{(Color online) The Abell~1835 grating spectrum upper limit on the decay rate and its approximate analytical expression given by of \eqnref{analytical}.}
\label{fig:fit}
\end{figure}

The constraints on the decay rate from the grating data of Abell~1835 can be approximated by an analytical expression for HEG and MEG independently. As seen in \figref{fit}, a second order polynomial describes the HEG constraints quite well, and a sixth order polynomial fits the MEG data:
\begin{eqnarray} \label{eqn:analytical}
\nonumber \Gamma_{HEG} &=& -8.6\cdot10^{-28} + 1.7\cdot10^{-27}E_\gamma - 1.5\cdot10^{-28}E_\gamma^2 \\
\nonumber \Gamma_{MEG} &=& -2.9\cdot10^{-28} + 1.4\cdot10^{-27}E_\gamma + 9.7\cdot10^{-28}E_\gamma^2 \\
&& - 5.4\cdot10^{-28}E_\gamma^3 + 1.1\cdot10^{-28}E_\gamma^4 \\
\nonumber && - 1.3\cdot10^{-29}E_\gamma^5 + 6.4\cdot10^{-31}E_\gamma^6 \, .
\end{eqnarray}

\section{Constraining Mass and Mixing Angle}
By regarding all decay branches possible through oscillations, the mean lifetime of a sterile Dirac neutrino of mass, $m_s$, has been determined to be \cite{barger:1995,dolgov:2000}:
\begin{equation}\label{eqn:tau}
\tau = \frac{1}{\Gamma _{tot}}=\frac{f(m_s)\cdot 10^{20}}{\left(m_s/\textrm{kev} \right)^5 \sin^2(2\theta) }\textrm{sec}^{-1} \, ,
\end{equation}
where $\Gamma_{tot}$ is the total decay rate. $f(m_s)$ takes into account the open decay channels so that for $m_s < 1 \MeV$, where only the neutrino channel is open, $f(m_s)=0.86$. For Majorana neutrinos, which we will be considering below, $f(m_s)$ is half the value for Dirac neutrinos. The branching ratio of the radiative decay is $\Gamma_\gamma/\Gamma_{tot}=27\alpha/8\pi \approx 1/128$ \cite{barger:1995}. This can be combined with
\eqnref{maxdecay} to give:
\begin{eqnarray}
\sin^2(2\theta) &\lesssim& 7\times 10^{17}\textrm{sec}^{-1} \left(\frac{F_{det}}{\ergcms}\right) \left(\frac{m_s}{\keV}\right)^{-5} \\
\nonumber && \times \left[\frac{(M_{fov}/\Ms)}{(D_L/\Mpc)^2} + \frac{(M_{halo}/\Ms)}{(D_{halo}/\Mpc)^2}\right]^{-1} \, .
\end{eqnarray}

\begin{figure}
\center
\epsfxsize=8.6cm
\epsfysize=6.1cm
\epsffile{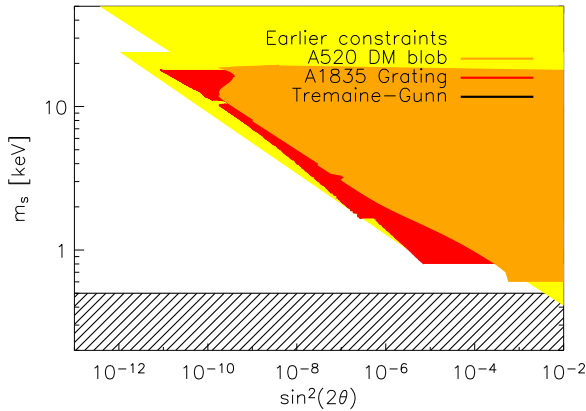}
\caption{(Color online) The observational constraints from the grating spectrum of Abell~1835 (red) and the dark matter blob of A520 (orange) together with the Tremaine-Gunn limit (black, \cite{tremaine}) and earlier X-ray constraints (yellow, \cite{boyarsky:2005a,boyarsky:2006a,boyarsky:2006b,watson:2006,blanksky}).}
\label{fig:sin-m}
\end{figure}

The observational constraints in the $\sin^2(2\theta)-m_s$ parameter space are shown in \figref{sin-m} for the grating spectrum of Abell~1835 (red) and the dark matter blob of A520 (orange) together with the Tremaine-Gunn limit (black, \cite{tremaine}) and earlier X-ray constraints (yellow, \cite{boyarsky:2005a,boyarsky:2006a,boyarsky:2006b,watson:2006,blanksky}). It is seen that even though no optimization of the field of view can be performed for the grating data, the improvement of instrumental spectral resolution leads to superior constraints. The grating constraints are very robust as they have been derived from the total amount of received X-ray flux without subtraction of any background contributions. The earlier constraints are derived by less robust and more model dependent methods.

\section{Conclusions}
A very general constraint on the decay rate for all dark matter particle candidates with a two-body radiative decay in the X-ray range has been derived. We have analyzed grating spectra of the galaxy cluster Abell~1835, and a spectrum obtained through direct imaging of the almost pure dark matter blob in the galaxy cluster Abell~520.
These provide the strongest constraints on radiatively decaying dark matter derived in a very conservative way from cosmic X-ray data. The mass and mixing angle can be constrained in the specific case of sterile neutrinos, leaving a low mass ($m_s \lesssim 10$~keV) and low mixing angle ($sin^2(2\theta) \lesssim 10^{-6}$) window open. However, combined with other methods \cite{seljak06,viel05}, the Dodelson-Widrow scenario for sterile neutrino production could soon be ruled out. Other mechanisms for production of sterile neutrinos, for example, via inflaton
decays \cite{shaposhnikov:2006}, would satisfy all the constraints.

The obtained constraints can be improved significantly by improving the signal to noise ratio (optimization of field of view) and by improving the instrumental spectral resolution. This could be realized through X-ray grating observations of dark matter dense regions or the Milky Way halo (blank sky fields).

\acknowledgments The Dark Cosmology Centre is funded by the Danish National Research Foundation. KP acknowledges support from Instrument Center for Danish Astrophysics. We thank M.~Shaposhnikov for pleasant and useful discussions, and A.~Boyarsky and O.~Ruchayskiy for useful comments. 


\end{document}